\documentclass[showpacs,preprint, nofootinbib]{revtex4-1}
\usepackage{amssymb}

\newcommand{\be}{\begin{equation}}
\newcommand{\en}{\end{equation}}
\newcommand{\bea}{\begin{eqnarray}}
\newcommand{\ena}{\end{eqnarray}}
\input{tcilatex}

\begin{document}

\title{Hawking radiation from z=3 and z=1-Lifshitz black holes}
\author{ Samuel Lepe}
\email{slepe@ucv.cl}
\author{Bruno Merello}
\email{bruno.merello.e@mail.pucv.cl}
\date{\today}

\begin{abstract}
The Hawking radiation considered as a tunneling process, by using a
Hamilton-Jacobi prescription, is discussed for both $z=3$ and $z=1$-Lifshitz
black holes. We have found that the tunneling rate (which is not thermal but
related to the change of entropy) for the\textbf{\ }$z=3$-Lifshitz black
hole (which does not satisfy the Area/4-law) does not yield (give us) the
expected tunneling rate: $\Gamma \sim \exp \left( \Delta S\right) $, where $%
\Delta S$\ is the change of black hole entropy, if we compare with the $z=1$%
-Lifshitz black hole (BTZ black hole, which satisfies the Area/4-law).
\end{abstract}

\maketitle

\affiliation{\it Instituto de F\'{\i}sica, Pontificia Universidad
de Cat\'{o}lica de Valpara\'{\i}so, Casilla 4950, Valpara\'{\i}so,
Chile}

\section{Introduction}

The second law of thermodynamics was initially put forth for a system
including black holes by Bekenstein [1]. It states that sum of one quarter
of the area black hole's event horizon plus the entropy of ordinary matter
outside never decrease with time all process. The thermodynamics properties
of these objects are based on its elusive quantum description (as everything
seems to indicate). Specifically, the problem of emission of particles and
its relationship to the second law of thermodynamics, i. e., the entropy.
Black holes can emit particles from the event horizon known as Hawking
radiation [2]. Hence, the classical notion is no longer that a black hole is
not as black,\textbf{\ }or in other words, the black hole radiation appears
as a quantum process (as everything seems to indicate). Whilst the nature of
Hawking radiation is not yet fully understood [3], this fact has established
an important antecedent when exploring the evolution of a black hole and its
thermodynamic behaviour taking into account the underlying holographic idea
about the crucial relation between entropy and area in a gravitacional
system [4]. In this work we study the Hawking radiation from a couple of
Lifshitz black holes ($z=3$ and $z=1$ cases) by following an idea proposed
in [5]: the Hawking radiation from a black hole described as a tunneling
process. The results for both cases are compared and discussed. The present%
\textbf{\ }paper is organized as follows. In Section 2, we review the $z=3$%
-Lifshitz black hole solution and its themodynamics. In Section 3, we
compute the tunneling rate (which is not thermal ) for this black hole by
following a prescription based on the relativistic Hamilton-Jacobi equation
and we compare the obtained result with the one\textbf{\ }corresponding to $%
z=1$-Lifshitz black hole (BTZ black hole). In Section 4 we discuss our
results. In this paper, the units G=%
%TCIMACRO{\U{127}}%
%BeginExpansion
h{\hskip-.2em}\llap{\protect\rule[1.1ex]{.325em}{.1ex}}{\hskip.2em}%
%EndExpansion
=c=1 are used.

\section{$z=3$\textbf{-Lifshitz black hole}}

The black hole solution is decribed by the following metric [6] 
\begin{equation}
ds^{2}=-\left( \frac{r^{2}}{l^{2}}\right) ^{3}\left(
1-r_{+}^{2}/r^{2}\right) dt^{2}+\frac{l^{2}}{r^{2}}\left(
1-r_{+}^{2}/r^{2}\right) ^{-1}dr^{2}+r^{2}d\varphi ^{2},  \tag{1}
\end{equation}%
where the coordinates are defined by $-\infty \leq t\leq \infty $, $r\geqq 0$
and $0\leq \varphi \leqslant 2\pi $, and $r_{+}=l\sqrt{M}$ is the black hole
event horizon.\textbf{\ }Here $M$ is an integration constant related to the
black hole mass and\textbf{\ }$l$ is the curvature radius of\ the Lifshitz
spacetimes related to the cosmological constant through $\Lambda =-13/2l^{2}$
. The thermodynamical quantities of this black hole [7], i.e.\textbf{\ }%
temperature, entropy and mass are, respectively, 
\begin{equation}
T=\frac{r_{+}^{3}}{2\pi l^{4}}\text{ \ },\text{ \ }S=2\pi r_{+}\text{\ \ },%
\text{\ \ }\hat{M}=\frac{r_{+}^{4}}{4l^{4}}=\left( \frac{M}{2}\right) ^{2}, 
\tag{2}
\end{equation}%
such that $TdS=d$\ $\hat{M}$; meaning that the first law is satisfied but
not the Area/4-law for the entropy: $S=2\pi r_{+}\neq \pi r_{+}/2=Area/4$.\
This point will be very important\ at time to calculate the tunneling rate
for this black hole and the subsequent comparison with the $z=1$-Lifshitz
black hole (BTZ black hole). For $z\neq 1,3$ it is not clear yet how to
compute conserved quantities for\textbf{\ }asymptotic Lifshitz black holes
and this fact make\textbf{\ }difficulties\textbf{\ }if we want to have a
consistent thermodynamic: the thermodynamical laws should be verified for
this type of black holes. Whether or not the Area/4-law is a universal law
which must satisfy all black holes, independent of the dimensionality of the
spacetime in which they live, is still an open question. We conclude this
Section with an example (there are others) which does not appear to be of any%
\textbf{\ }help\textbf{\ }in finding an answer to the previous question: in
a scattering process of scalar fields over a $z=3$-Lifshitz black hole [8],\
the absorption cross-section is proportional to the entropy independent
whether or not the Area/4-law is fulfilled.

\section{ The relativistic Hamilton-Jacobi equation}

We use a semi-classical method proposed in Ref. [9] in order to\textbf{\ }%
conceive\textbf{\ }the Hawking radiation as a tunneling effect. The idea is
to consider a scalar particle moving in the background of the black hole
where the particle self-gravitation is neglected. So, the classical action $%
I $ of the particle satisfies the relativistic Hamilton-Jacobi equation 
\begin{equation}
g^{\mu \nu }\left( \frac{\partial I}{\partial x^{\mu }}\right) \left( \frac{%
\partial I}{\partial x^{\nu }}\right) +m^{2}=0,  \tag{3}
\end{equation}%
where $m$ and $g^{\mu \nu }$ are the mass of the particle and the inverse
metric tensor derived from (1) and $x^{\mu }=\left( t,r,\varphi \right) $.
Near the horizon 
\begin{equation}
\left( \frac{r^{4}}{l^{6}}\Delta \left( r\right) \right) _{r\longrightarrow
r_{+}}\longrightarrow \frac{2r_{+}^{5}}{l^{6}}\left( r-r_{+}\right) \text{ \
\ },\text{ \ \ }\left( \frac{l^{2}}{\Delta \left( r\right) }\right)
_{r\longrightarrow r_{+}}\longrightarrow \frac{l^{2}}{2r_{+}\left(
r-r_{+}\right) },  \tag{4}
\end{equation}
\textbf{such} that in this case, where the non-null inverse metric elements
are 
\begin{equation}
g^{tt}\left( r\right) =-\frac{l^{6}}{2r_{+}^{5}}\frac{1}{\left(
r-r_{+}\right) }\text{ \ \ },\text{ \ \ }g^{rr}\left( r\right) =\frac{2r_{+}%
}{l^{2}}\left( r-r_{+}\right) \text{ \ \ },\text{ \ \ \ }g^{\varphi \varphi
}\left( r\right) =\frac{1}{r_{+}^{2}}.  \tag{5}
\end{equation}

By replacing (5) in (3), and because there are two Killing vectors in the
present $z=3$-black hole, we do $I=-\omega t+R(r)+j\varphi $, where $\omega $
and $j$ are the energy and angular momentum of the particle respectively and 
$R\left( r\right) $ is\textbf{\ }the geometric content of the spacetime
under consideration, we have the following integral expresion for $R\left(
r\right) $%
\begin{equation}
R\left( r\right) =\pm \frac{l^{4}\omega }{2r_{+}^{3}}\int \frac{dr}{r-r_{+}}%
\sqrt{1-2\left( \frac{r_{+}^{2}}{\omega l^{3}}\right) ^{2}r_{+}\left( m^{2}+%
\frac{j^{2}}{r_{+}^{2}}\right) \left( r-r_{+}\right) },  \tag{6}
\end{equation}%
expression from which we rescue the imaginary part (classically forbidden
process) which is related to the Boltzmann factor for emision at the Hawking
temperature 
\begin{equation}
R\left( r\right) =\pm \frac{l^{4}\omega }{2r_{+}^{3}}\left( i\pi \right)
\Longrightarrow I=-\omega t\pm \frac{l^{4}\omega }{2r_{+}^{3}}\left( i\pi
\right) +j\varphi .  \tag{7}
\end{equation}%
The $\pm $ sign in (7) correspond to outgoing and ingoing particles,
respectively. Given that in the classical limit all is absorbed by the black
hole (with no reflection) we write $Im I$ corresponding to outgoing
particles 
\begin{equation}
\mathit{{Im}I=\frac{l^{4}\omega \pi }{2r_{+}^{3}}=\frac{\omega }{4T}=\frac{%
\pi l\omega }{2^{5/2}}\hat{M}^{-3/4},}  \tag{8}
\end{equation}%
where $r_{+}^{2}/l^{2}=2\sqrt{\hat{M}}$ ($\Re I$ corresponds to ingoing
particles, i. e., particles falling through the horizon of the black
hole).We note that $2\mathit{{Im}I=\omega T/2}$ and already this fact shows
the non-thermal nature of the emission; for thermal emission $2\mathit{{Im}%
I=\omega T}$. Now, when a particle with energy $\omega $ tunnels out, the
mass of the black hole changed into $\hat{M}\longrightarrow \hat{M}-\omega $
($l\sqrt{M}\longrightarrow l\sqrt{M-\omega }$, where $l\sqrt{M}$ is the
horizon before pair-creation and $l\sqrt{M-\omega }$ the horizon after
pair-creation).\textbf{\ }Therefore we put 
\begin{equation}
Im I\rightarrow -\frac{\pi l}{2^{5/2}}\int_{\hat{M}}^{\hat{M}-\omega
}d\left( \hat{M}-\omega \right) \left( \hat{M}-\omega \right) ^{-3/4}, 
\tag{9}
\end{equation}%
and we obtain 
\begin{equation}
Im I=-\frac{1}{4}\left[ \left( 1-\frac{\omega }{\hat{M}}\right) ^{1/4}-1%
\right] S=-\frac{1}{4}\Delta S,  \tag{10}
\end{equation}%
where $S=2\pi r_{+}$ is the $z=3$-Lifshitz black hole entropy. In the WKB
approximation the tunneling rate (tunneling probability for the classically
forbidden trajectory from inside to outside the horizon) is given by $\Gamma
\sim \exp \left( -2\mathit{{Im}I}\right) $ and for this black hole we find
Therefore, the tunneling rate $\Gamma \sim \exp \left( -2 Im I\right) $ for
this black hole is given by 
\begin{equation}
\Gamma \sim \exp \left( \frac{1}{2}\Delta S\right) ,  \tag{11}
\end{equation}%
\ and we note that (11) differs from the result $\Gamma \sim \exp \left(
\Delta S\right) $ which is valid for black holes that satisfy the
Area/4-law. If we repeat the scheme done before for the BTZ black hole ($z=1$%
-Lifshitz black hole) we find [10] 
\begin{equation}
\Gamma \sim \exp \left( \Delta S\right) ,  \tag{12}
\end{equation}%
where 
\begin{equation}
\Delta S=\left[ \left( 1-\frac{\omega }{M_{BTZ}}\right) ^{1/2}-1\right] S, 
\tag{13}
\end{equation}%
being $S=\pi r_{+}/2=2\pi r_{+}/4=Area/4$-$law$ and $r_{+}=l\sqrt{M}=2\sqrt{2%
}l\sqrt{M_{BTZ}}$, such that $dM_{BTZ}=TdS$ (first law) and $T=r_{+}/2\pi
l^{2}$ [6].

\section{Concluding remarks.}

We have computed the tunneling rate for both $z=3$-Lifshitz black hole,
which does not satisfy the Area/4-law and we find $\Gamma \sim \exp \left(
\Delta S/2\right) $, and $z=1$-Lifshitz black hole (BTZ), which satisfies
the Area/4-law and we found\textbf{\ } $\Gamma \sim \exp \left( \Delta
S\right) $. It is not yet clear how to compute conserved quantities in
asymptotic Lifshitz black holes and so, we can not visualize the scope of
our results if we are thinking in a complete\textbf{\ }thermodynamical
description of these type of black holes in the framework, for instance, of
new massive gravity. Nevertheless, we apologize that the difference in the
tunneling rates obtained can be a signal to discriminate between black holes
which satisfy the Area/4-law and those who do not. Finally, in the
literature we can find discussions about dependence on the type of
coordinates used to describe black holes and then calculate tunneling rates
(see [11] and references therein). However, we believe that the
thermodynamic properties of these, if they do, must be independent of that
choice, as it should be if we accept (assume?) that black holes are thermal
objects. 
\begin{acknowledgments}
This work was supported by Fondecyt Grant No. 1110076 (SL) and VRIEA-DI-PUCV
Grant No. 037.377/2014, Vice Rector\'{\i}a de Investigaci\'{o}n y Estudios
Avanzados, Pontificia Universidad Cat\'{o}lica de Valpara\'{\i}so (SL). We
thank U. Raff for a careful reading of the manuscript. Also we
acknowledge Dr. Joel Saavedra for helpful discussions.
\end{acknowledgments}


\begin{thebibliography}{99}
\bibitem{01} J. D. Bekenstein, Nuovo Cim. Lett \textbf{4}, 737 (1972), Phys.
Rev. D\textbf{7}, 2333 (1973) and Phys. Rev. D\textbf{9}, 3292 (1974).

\bibitem{1} S. W. Hawking, Nature \textbf{30}, 248 (1974), S. Hawking,
Commun. Math. Phys. \textbf{43} (1975) 199-220.

\bibitem{2} R. Sch\^{u}tzhold and W. G. Unruh, arXiv:1308.2159 [gr-qc].

\bibitem{3} K. Jensen and A. O`Bannon, Phys. Rev. D\textbf{88} (2013)
106006, A. Bhattacharyya, M. Sharma and A. Sinha, JHEP \textbf{1401} (2014)
021.

\bibitem{4} M. K. Parikh and F. Wilczek, Phys. Rev. Lett. \textbf{85} (2000)
5042-5045.

\bibitem{5} Eloy Ay\'{o}n-Beato et al, Phys. Rev. D\textbf{80} (2009) 104029.

\bibitem{6} Yun Soo Myung and Taeyoon Moon, Phys. Rev. D\textbf{86} (2012)
024006.

\bibitem{7} S. Lepe et al, Phys. Rev. D\textbf{86} (2012) 066008.

\bibitem{8} M. Angheben et al, JHEP \textbf{0505} (2005) 014. 

\bibitem{9} Xiaokai He and Wembiao Liu, Phys. Lett. B\textbf{653} (2007)
330-334.

\bibitem{11} Ren Jun, Jia Meng-Wen and Wei Yi-Huan, Chin. Phys. B\textbf{22}
(2013) 11, 110401.
\end{thebibliography}
\end{document}